\documentclass[aps,prb,twocolumn,showpacs]{revtex4}

\usepackage{latexsym}
\usepackage{graphicx}
\usepackage{times,psfrag,subfigure}
\usepackage{amsmath}
\usepackage{dcolumn}
\usepackage{color}
\usepackage{latexsym,amsmath,amssymb,bm,euscript}
\newcommand{\fss}{FeSc$_2$S$_4$}

\newcommand{\mc}{\mathcal}

\begin{document}

\title{Excitation spectrum and magnetic field effects in a quantum
  critical spin-orbital system: \fss}

\date{\today}

\author{Gang Chen}
\affiliation{Physics Department, 
University of California, Santa Barbara, CA 93106}
\author{Andreas P. Schnyder}
\affiliation{Kavli Institute for Theoretical Physics, 
University of California, Santa Barbara, CA 93106}
\author{Leon Balents}
\affiliation{Kavli Institute for Theoretical Physics,
University of California, Santa Barbara, CA 93106}

\begin{abstract}
  The orbitally degenerate A-site spinel compound \fss\ has been
  experimentally identified as a ``spin-orbital liquid'', with strong
  fluctuations of both spins and orbitals.  Assuming that the second
  neighbor spin exchange $J_2$ is the dominant one, we argued in a
  recent theoretical study [Chen \textit{et al.}, Phys.\ Rev.\ Lett.\
  \textbf{102}, 096406 (2009)] that \fss\ is in a local ``spin-orbital
  singlet'' state driven by spin orbit coupling, close to a quantum
  critical point, which separates the ``spin-orbital singlet'' phase
  from a magnetically and orbitally ordered phase.  In this paper, we
  refine further and develop this theory of \fss.   
  First, we show that inclusion of a small first neighbor
  exchange $J_1$ induces a narrow region of incommensurate phase near
  the quantum critical point.  Next, we derive
  the phase diagram in the presence of an external magnetic field $B$,
  and show that the latter suppresses the ordered phase.  Lastly, we
  compute the field dependent dynamical magnetic susceptibility
  $\chi({\bf k},\omega;B)$, from which we extract a variety of physical
  quantities.  Comparison with and suggestions for experiment are discussed. 
\end{abstract}
\date{\today}

\pacs{71.70.Ej,71.70.Gm,75.10.-b,75.40.-s}


\maketitle


\section{Introduction}
\label{sec:sec1}

Among all the magnetic spinels,\cite{tristan:prb05,krimmel:pb,suzuki:jpcm,kalvius:physicaB09,loidl:ns,fritsch:prl04,loidl:nsnmr,krimmel:prb1,giri:prb,loidl:nmr,kalvius:physicaB06,fichtl:05,krimmel:PRB09}
 the A-site spinel \fss\ is particularly
intriguing in that its frustration
parameter\cite{ramirez:review94,ramirez:hndbk} $f \gtrsim 1000$ is one
of the largest ever reported.\cite{fritsch:prl04,loidl:nsnmr,loidl:ns}
Indeed, even though the material clearly exhibits well-formed local
moments interacting with a characteristic energy given by the
Curie-Weiss temperature $| \Theta_{CW} |=45$K, no sign of magnetic
ordering has been found down to the lowest measurable temperature of
$50$mK.  Moreover, \fss\ is interesting because it has not only spin but
also orbital degeneracy.  The Fe$^{2+}$ ion at the A-site is in a 3$d^6$
configuration, whose five-fold degeneracy is split by the tetrahedral
crystal field
into a lower $e_g$ doublet and an upper $t_{2g}$ triplet. 
The six electrons in the 3$d$ shell are Hund's rule coupled, yielding
a high spin configuration with $S=2$ and a two-fold
orbital degeneracy due to a hole in the lower $e_g$ doublet.
Besides the five-fold spin degeneracy and the lattice degrees of freedom,
the orbital degeneracy gives an additional contribution to
the specific heat. This has been confirmed experimentally.\cite{fritsch:prl04,loidl:nsnmr}

Commonly, orbital degeneracies are relieved by a
Jahn-Teller type structural distortion, that leads to orbital order at
low temperatures.  However, in \fss\ no such distortion has been
observed.  Hence, both the spins and orbitals remain frustrated and
continue to fluctuate down to the lowest measured temperature, a
situation for which the term ``spin orbital liquid'' (SOL) was coined.
A SOL state was also suggested for
LiNiO$_2$. \cite{feiner:prl97,kitaoka:jpsj98} But this proposal has been
questioned recently in Ref.~\onlinecite{mostovoy:prl}, where it was
argued that the unusual behavior of LiNiO$_2$ is due to disorder
effects.  Therefore \fss\ remains as the best candidate for a SOL.

\begin{figure}[tp]
\includegraphics[width=6.0cm]{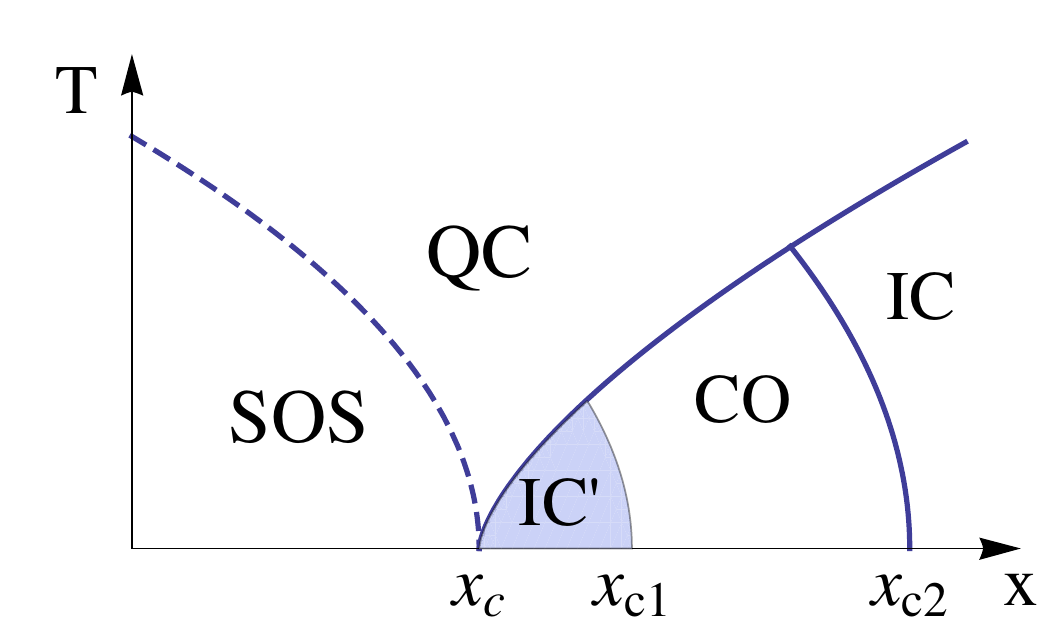}
\caption{(color online).  Schematic phase diagram as a function of
 temperature $T$ and $x$, the ratio of exchange to spin-orbit interaction.
A quantum critical point (QCP) at $x_c$ separates 
the spin-orbital-singlet (SOS) state from the ordered phase.
Within the order phase, at $x_{c2}$, there is a commensurate-incommensurate
phase transition. ``QC'' denotes the quantum critical regime,
``CO'' the commensurate antiferromagnet with orbital order, 
and ``IC''  the incommensurate spin and orbital order. 
Inclusion of a small first neighbor exchange $J_1$ 
induces a second region of incommensurate phase (shaded area).%
}
\label{fig:ph1}
\end{figure}
 
In a undistorted lattice, exchange interactions or spin-orbit coupling
can split the orbital and spin degeneracies of a single Fe$^{2+}$
ion.  To study these possibilities we introduced in
Ref.~\onlinecite{chen:prl} a model that contains
a ``Kugel-Khomskii''-type \cite{kk:82} spin-orbital exchange interaction
as well as the lowest-order symmetry-allowed atomic spin-orbit coupling.
The exchange interactions favor spin and orbital order, whereas
the on-site spin-orbit coupling leads to the formation of a local
``spin-orbital singlet'' (SOS).  In \fss\ there is strong competition
between these two interactions. We argued in Ref.~\onlinecite{chen:prl}
that \fss\ is in the SOS state, close to a quantum critical point (QCP)
that separates the SOS phase from the ordered phase (see
Fig.~\ref{fig:ph1}).  This QCP seems to be rather analogous to that
appearing in spin-dimer materials \cite{nikuni1971bec,jaimePRL04} or bilayer Heisenberg 
models,~\cite{hida1992}
with the two orbital states playing the roles of the two members of a spin dimer,
or the two bilayer states.
However, we will discover at least
one surprising difference below.  

Previously, we have argued that, to a first approximation, \fss\ can be
described by a simplified ``$J_2 - \lambda$ model'',\cite{chen:prl}
which only contains on-site spin-orbit and next-nearest neighbor (NNN)
spin exchange interactions.  Within this approximation the two fcc
sublattices of the A-site diamond lattice decouple completely.  The aim
of the present paper is to refine the model of
Ref.~\onlinecite{chen:prl} and to calculate more detailed physical
properties for comparison with experiments.  

\begin{figure}[tp]
\includegraphics[width=0.35\textwidth]{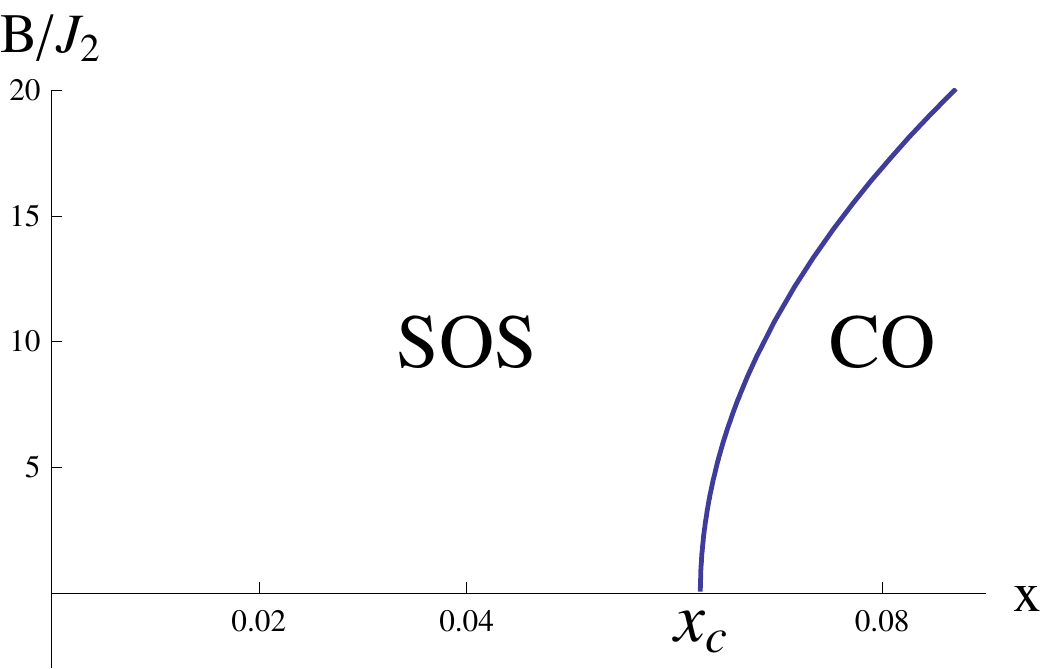}
\caption{(color online).  Phase diagram of the $J_2-\lambda$ model, 
Eq.~(\ref{eq:min_mod}), as a
function of magnetic field $B$ and coupling ration $x$.
In zero magnetic field the phase transition between the spin-orbital singlet (SOS) phase
and the commensurate ordered (CO) state occurs at $x_c=1/16$.}
\label{fig:ph2}
\end{figure}

We summarize the results here.  We first consider
the effects of weak intersublattice nearest-neighbor (NN) exchange
$J_1$.  We find that it induces a narrow range of {\sl incommensurate}
magnetically ordered phase in the vicinity of the QCP, on the
ordered side (see Fig.~\ref{fig:ph1}).  Next, we consider the effects of an external magnetic
field $B$ on the QCP in the $J_2 - \lambda$ model.  Quite surprisingly,
we find that the magnetic field {\sl destroys} the spin order and the
QCP shifts from $x_c = 1/16$ to a larger value (see Fig.~\ref{fig:ph2}).
This is exactly the {\sl opposite} trend to that observed in the
spin-dimer and bilayer models.  Within mean field theory we calculate
both the uniform and staggered magnetization as a function of $B$.
Finally, we use the random phase approximation (RPA) to compute the
field-dependent dynamical spin susceptibility $\textrm{Im} \chi (
\bm{k}, \omega, B )$ in the SOS phase.  From an analysis of the pole
structure of $\chi ( \bm{k}, \omega )$ we obtain the dispersion of the
low-energy collective modes.  Consistent with the effect of the magnetic
field on the phase diagram, we find that the gap in the SOS phase
increases with $B$.  

The remainder of the paper is organized as follows.  In
Sec.~\ref{sec:model-definition}, we define the model Hamiltonian and
summarize the principle results of Ref.~\onlinecite{chen:prl}.  The weak
inter-sublattice interaction is introduced in Sec.~\ref{sec:sec3}, where
we derive the resulting changes to the phase diagram.  In
Sec.~\ref{sec:sec2}, we return to the minimal NNN model, but include the
effects of an external magnetic field.  The inelastic structure factor
(dynamical spin susceptibility) is calculated by the RPA approximation
in Sec.~\ref{sec:subsec22}.  We conclude with a summary and discussion
in Sec.~\ref{sec:sec5}.  Some technicalities are given in the appendix.

\section{Model definition}
\label{sec:model-definition}

The Hamiltonian contains two terms:
an exchange interaction $\mc{H}_{\text{ex}}$ and an atomic spin-orbit coupling
$\mc{H}^i_0$,
\begin{subequations} \label{eq:ham_parent}
\begin{eqnarray} \label{eq:ham_parent a}
\mc{H}  
=
  \mc{H}_{\text{ex}} + \sum_i \mc{H}_0^i .
\end{eqnarray}
The term $\mc{H}_{\text{ex}}$ describes spin and orbital exchange interactions
as well as couplings between spin and orbital degrees of freedom.
Using microscopic considerations and symmetry constraints
one can show that $\mc{H}_{\text{ex}}$ is of the form~\cite{chen:prl}
\begin{eqnarray}  \label{eq:ham_parent b}
\mc{H}_{\text{ex}}  
=
\tfrac{1}{2}\sum_{ij} 
\left[
J_{ij}\ {\bf S}_i \cdot {\bf S}_j 
+ 2K_{ij} {\bf T}_i \cdot {\bf T}_j (4 + {\bf S}_i \cdot {\bf S}_j )
\right]  ,
\end{eqnarray}
where ${\bf S}_i$ is the $S=2$ spin operator at site $i$.
The orbital degrees of freedom are
described by $T=1/2$ 
pseudospin operators ${\bf T}_i$
that act on 
the $ x^2 -y^2  $ and $  3 z^2 -r^2 $ orbitals in the $e_g$ subspace.
An analysis of the exchange paths linking two A-sites 
shows that both first and
second neighbor exchange paths are of comparable length and
have similar multiplicity.~\cite{roth:JdPhys,loidl:nsnmr}
This suggests that 
there is a substantial NNN interaction
and it is therefore necessary to keep in  Eq.~(\ref{eq:ham_parent b}) the sum 
over both the NN and NNN sites.
For convenience we set  $J_{ij}=J_1$ or $J_2$ when $ij$ are first and
second neighbor sites, respectively (and similarly for $K_{ij}$).

The second term in Eq.~(\ref{eq:ham_parent a}), $\mathcal{H}^i_0$,
is the on-site spin-orbit coupling which arises from second
order perturbation theory~\cite{vallin:prb}
\begin{eqnarray}  \label{eq:ham_parent c}
{\mathcal H}_0^i 
=
 -\tfrac{\lambda}{3} 
\left\{ 
 \sqrt{3}
  T_i^x 
[ (S_i^x)^2-(S_i^y)^2 ] + T_i^z  [ 3(S_i^z)^2-{\bf S}_i^2 ] 
 \right\} ,
\nonumber\\
\end{eqnarray}
\end{subequations}
where the coefficient $\lambda$ is estimated from microscopic
atomic calculations to be
$\lambda \simeq 6 \lambda_0^2 / \Delta_{\textrm{te}}$.
Here, $\lambda_0$ denotes the atomic spin-orbit interaction  and  $\Delta_{\textrm{te}}$ 
is the energy separation between
the $e_g$ and $t_{2g}$ states. 
Note, that $\mathcal{H}^i_0$ immediately leads to a splitting of 
the ionic degeneracies because the  ground
state of $\mathcal{H}^i_0$ is the non-degenerate spin-orbital singlet
\begin{eqnarray}
&&
\tfrac{1}{\sqrt{2}} 
\left| x^2 -y^2 \right\rangle
\left| S^z \vspace{-0.2cm} = 0 \right\rangle
+
\\
&& \qquad \quad
\tfrac{1}{2} 
\left| 3 z^2 -r^2 \right\rangle
\Big[
\left| S^z=+2 \right\rangle
+
\left| S^z=-2 \right\rangle
\Big] .
\nonumber
\end{eqnarray}
Whether \fss\ is in such a SOS phase depends
on the ratio between exchange  and on-site spin-orbit interactions
\begin{equation} \label{ratio x}
x  \equiv {\rm max} \left\{ J_{1}, J_{2}, K_{1}, K_{2}  \right\}/\lambda .
\end{equation}

We note that we have {\sl not} included spin-orbit effects such as
Dzyaloshinskii-Moriya interactions in the exchange Hamiltonian,
Eq.~(\ref{eq:ham_parent b}).  This may be surprising since the on-site
spin-orbit coupling in Eq.~(\ref{eq:ham_parent c}) plays a crucial role
in our analysis.  However, we expect that the spin-orbit corrections to
the exchange are smaller than the leading order exchange couplings by a
factor of order $\lambda_0/\Delta_{te}$. 
This makes them subdominant both to the isotropic exchange couplings in
Eq.~(\ref{eq:ham_parent b}) and the on-site spin-orbit interaction in
Eq.~(\ref{eq:ham_parent c}), which are comparable in \fss\ (this indeed
defines the location of the QCP -- see below).

From a comparison with experiments we showed in
Ref.~\onlinecite{chen:prl} that $J_2$ is antiferromagnetic and the
largest among all the exchange coupling constants, whereas $\lambda$ is
competitive, but slightly larger than the NNN spin exchange interaction
$J_2$.  This observation leads us to consider a ``minimal'' version of
model (\ref{eq:ham_parent}), which only includes the NNN spin exchange
$J_2$ and the onsite spin-orbit interaction $\lambda$ (i.e.,
$J_1=K_1=K_2=0$).  Within this $J_2-\lambda$ model we demonstrated that
\fss\ is in the SOS phase close to the QCP of Fig.~\ref{fig:ph1}.

The full phase diagram of  Hamiltonian (\ref{eq:ham_parent}) as a function of temperature and ratio $x$, Eq.~(\ref{ratio x}),
 includes a commensurate-incommensurate transition within the ordered phase
(see Fig.~\ref{fig:ph1}). Deep in the ordered phase when the exchange interactions are dominant  ($x \gg 1$)
the incommensurate spin and orbital order (IC) is generally favored by the exchange Hamiltonian
\eqref{eq:ham_parent a}.
 With the inclusion of weak spin-orbit interaction, Eq.~\eqref{eq:ham_parent b},
 i.e., with decreasing $x$,
 the spin and orbital order becomes commensurate 
 with the spins 
 and orbitals both forming a spiral with wavevectors
 ${\bf p} = 2\pi (1,0,0)$ and  ${\bf q} = (0,0,0)$, respectively.

\section{Effects of weak inter-sublattice spin exchange interaction}
\label{sec:sec3}

As mentioned in the introduction, the inclusion of  a small NN interaction $J_1$
induces a narrow region of incommensurate order near the QCP (shaded area in Fig.~\ref{fig:ph1}).
In this section we give a derivation of this results  using a Landau expansion of the effective action.


\begin{widetext}
 First, we note that the Hamiltonian in
  Eq.~\eqref{eq:min_mod} has independent cubic ``internal'' spin
  symmetry and cubic ``external'' space group symmetry.  We therefore
  have the symmetry allowed free energy for two decoupled fcc
  sublattices \cite{chen:prl} near the QCP,
  \begin{eqnarray}
    \label{eq:3}
    f_0 (\{\psi\}) & = & \sum_{\mu=A,B;a}  \Big[ v_1^2
    |\partial_a{\boldsymbol\psi}_{\mu,a}|^2 + v_2^2 \sum_{b \neq a} |\partial_b{\boldsymbol\psi}_{\mu,a}|^2 
    + r |{\boldsymbol\psi}_{\mu,a}|^2 \Big] + g_1 \sum_{\mu,a}
   \big(     |\boldsymbol{\psi}_{\mu,a}|^2 \big)^2 + g_2 \sum_{\mu,a,b}
    (\psi_{\mu,a}^b)^4 
     \\
    & + & \sum_\mu {\rm Sym}\big[
    g_3 (\psi_{\mu,1}^x)^2(\psi_{\mu,2}^x)^2 
    + g_4 (\psi_{\mu,1}^x)^2(\psi_{\mu,2}^y)^2 + g_5 \psi_{\mu,1}^x \psi_{\mu,1}^y \psi_{\mu,2}^x \psi_{\mu,2}^y
    \big] \Big], \nonumber
    \label{eq:free_energy1}
  \end{eqnarray}
  where the order parameters $\psi_{\mu,a}$ are the (real) staggered
  magnetizations introduced by \begin{eqnarray} \langle {\bf S}_i \rangle = \left\{
    \begin{array}{ll}
      \sum\limits_{a=x,y,z} {\boldsymbol \psi}_{A,a}(-1)^{2 x_i^a} & i \in \text{A sublattice} 
      \vspace{0.2cm}   \\
      \sum\limits_{a=x,y,z} {\boldsymbol \psi}_{B,a}(-1)^{2 x_i^a} & i \in \text{B sublattice} 
    \end{array}
  \right.  \;, \end{eqnarray} 
  and $x_i^a$ are
the canonical half-integer coordinates of the fcc lattice with 
the cubic supercell having unit length.
For our convenience we choose the same set of fcc
  coordinates for the two sublattices. 
  In Eq.~\eqref{eq:free_energy1}, ``${\rm Sym}$''
  indicates symmetrization with respect to both wavevector (lower) and
  spin (upper) indices.
\end{widetext}

We now introduce weak inter-sublattice interactions. As a result, extra
terms which couple two fcc sublattices will appear in
Eq.~\eqref{eq:free_energy1}.  By the symmetry analysis, we obtain the
inter-sublattice terms
\begin{eqnarray}
  f_{\text{int}} \left(\{\psi\} \right) & = & \sum_a \gamma \
  \boldsymbol{\psi}_{A,a} \partial_a \boldsymbol{\psi}_{B,a}  
   - \eta \left(\boldsymbol{\psi}_{A,a} \cdot
    \boldsymbol{\psi}_{B,a}\right)^2,
  \label{eq:free_energy2}
  \nonumber\\
\end{eqnarray}
where we have kept the leading quadratic term (others have more
derivatives) and the most important quartic term, which is when
inter-sublattice couplings first appear without derivatives.  A number
of other quartic couplings involving the two sublattices are also
allowed, but do not play any important role in what follows.

We now proceed to analyze the Landau theory of
Eqs.~(\ref{eq:3}) and (\ref{eq:free_energy2}).  First consider the behavior on
approaching the QCP from the SOS phase.  The first instability of the
SOS phase is signalled by the quadratic part of the action alone,
i.e. the vanishing of the lowest eigenvalue of the associated quadratic
form.  Due to the linear derivative term, the unstable eigenvectors are non-constant
fields of the form
\begin{eqnarray}
  \label{eq:1}
  \psi^b_{A, a} & = & \psi_a^b \cos (\delta \, x_a +
  \theta_{a,b}),  \nonumber\\
  \psi^b_{B, a} & = & \psi_a^b \sin (\delta \, x_a + \theta_{a,b}), 
\end{eqnarray}
with $\delta = \gamma/(4v_1^2)$, and $\psi_a^b, \theta_{a,b}$
arbitrary constants. There is one linearly independent unstable
eigenvector for each $a$ and $b$.  The particular form of
superposition which is favored in the ordered state is determined by the
quartic terms.  We expect on physical grounds that the ordered states
will be of spiral type, with approximately constant {\sl magnitude} of
spin expectation values, and with a ``single $q$'' structure.  The
latter condition means that $\boldsymbol{\psi}_{A,a}$ is non-zero only
for a single $a=x,y$ {\sl or} $z$.  A sufficient condition for a
single $q$ structure to be favored is that $g_3,g_4,g_5>0$ (though this
condition can be relaxed somewhat).  
In this case, we have
\begin{eqnarray}
\label{eq:5}
  \boldsymbol{\psi}_{A, a} & = & \psi_0 {\rm Re} \left[( {\bf\hat e}_1 +
    i{\bf \hat e}_2 ) e^{i (\delta \, x_a + \theta)}\right], \\
  \boldsymbol{\psi}_{B, a} & = & \psi_0 {\rm Re} \left[( -{\bf\hat e}_2 +
    i{\bf \hat e}_1 ) e^{i (\delta  \, x_a + \theta)}\right], \nonumber
\end{eqnarray}
This corresponds to an incommensurate spiral state of spins with the
wavevector
\begin{equation}
{\bf p} = (2\pi \pm \delta ) (1,0,0)  
\;.
\end{equation} 
The unit vectors ${\bf \hat e}_{1/2}$ define the plane in which the
spins rotate.  At the quadratic level, this plane is arbitrary, but it
will be selected by the quartic terms in the free energy.  Presuming
that the system has axial cubic anisotropy (preferring spins aligned
with the $x$,$y$ or $z$ axes), a (100) or symmetry-related plane will be
chosen.  This is controlled in the free energy by the coefficient $g_2$,
which should be negative to mimic axial cubic anisotropy.  

Now consider the evolution of the spin configuration as the system
becomes more strongly ordered.  As $|{\boldsymbol\psi}|$ increases, we
expect the quartic terms in the free energy to become more important,
favoring commensurate states in which the spins are aligned with the
principle axes.  To analyze how this occurs, we presume that the spins remain in a
``single $q$'' structure (a spiral rather than a more exotic ``spin
lattice''), so that $\boldsymbol{\psi}_{\mu,a}$ is non-zero only for one
$a$.  Furthermore, we assume that the spins remain in a single (100)
plane.  Up to symmetry-related choices we take $a=x=1$, and ${\bf \hat
  e}_1 = {\bf \hat x}, {\bf\hat e}_2 ={\bf\hat y}$ in Eq.~(\ref{eq:1}).
Finally, we presume that the fields depend only upon $x$, as can be
easily verified is true for the minimum free energy configurations.  

Inserting this into Eqs.\ (\ref{eq:3}) and (\ref{eq:free_energy2}), and assuming
$\psi_0$ is constant, we obtain the reduced free energy
density, neglecting an additive constant
\begin{eqnarray}
  \label{eq:2}
  f & = & \frac{\kappa}{2} (\partial_x \theta)^2  + \sigma \cos
  4(\theta + \delta x),
\end{eqnarray}
with $\kappa= 4 v_1^2 \psi_0^2$, $\sigma =  g_2 \psi_0^4/2$.
This is a standard sine-Gordon model, with incommensuration $\delta$.
It describes a competition between a commensurate state in which $\theta
+ \delta x$ is constant, and an incommensurate one in which
$\partial_x\theta< \delta$ on average.  The transition between the two
states is known as a Commensurate-Incommensurate-Transition (CIT), and
its location is determined by the condition that the energy of a single
domain wall excitation of the commensurate state (a ``soliton'')
vanishes.  In this way we can precisely determine the location of the
CIT.  We find that the CIT occurs for~\cite{ChaikinLubensky}
\begin{equation}
  \label{eq:4}
  \delta_c = \frac{2}{\pi} \sqrt{\frac{|\sigma|}{\kappa}}.
\end{equation}
When $\delta<\delta_c$, the system is in a commensurate state.

We must now translate this condition back to obtain the critical point
in terms of microscopic parameters $x$, $x_c$, and $J_1/J_2$.  In
Appendix~\ref{sec:appendix}, we determine the necessary coefficients in
Eqs.\ \eqref{eq:3} and \eqref{eq:free_energy2} ($v_1^2,r,g_1,g_2,\gamma$) by
deriving the effective action by standard Hubbard-Stratonovich methods
from the microscopic spin-orbital Hamiltonian. The results are given in
Eqs.~\eqref{eq:coupling1} and \eqref{eq:coupling2}.  In addition, we require
the amplitude $\psi_0$ to minimize the free energy. 
With this we can calculate $\kappa$ and $\sigma$, and hence
solve Eq.~\eqref{eq:4} for the location of the CIT.  To get the
amplitude, we recognize that at the CIT the solution is commensurate,
i.e. has the form of Eq.~\eqref{eq:5} with $\delta=\theta=0$.  Therefore, we
may simply evaluate the free energy in Eq.~\eqref{eq:3} using this form
of the order parameter, and minimize over $\psi_0$.  The result is that
\begin{equation}
  \label{eq:6}
    \psi_0^2 = \frac{-r}{2(g_1+g_2)},
\end{equation}
where at the CIT of course $r<0$.  Combining the above results, we
obtain the CIT transition point
\begin{equation}
  \label{eq:7}
    x_{c1} = \frac{1}{16}\left[1+ \frac{ \pi^2}{4} \left( \frac{J_1}{J_2}\right)^2\right],
\end{equation}
for $J_1\ll J_2$.  As expected, for small $J_1$, the incommensurate
phase studied here is narrow and located only near the QCP (see Fig.~\ref{fig:ph1}).  As
discussed in Ref.~\onlinecite{chen:prl}, a different incommensurate phase
arises for much larger $J_2$, when the spin-orbit interaction plays a
minimal role.  This transition to the second incommensurate phase occurs
at $x_{c2} \approx 0.61 (J_2/J_1)^2$, far from the QCP.  

\section{$J_2-\lambda$ model in a magnetic field}

\label{sec:sec2}

In this section we study the minimal $J_2-\lambda$ model  
in an external magnetic field $B$ both in the CO and the
SOS phase (see Fig.~\ref{fig:ph1}). For definitiveness we assume
that the spins in the CO phase  
align themselves along the $x$ axis, whereas the magnetic field
is applied along the $z$ axis. 
We use mean field theory to calculate the uniform 
and staggered magnetizations and derive therefrom
the magnetic phase diagram of the $J_2-\lambda$ model.

\label{sec:subsec21}

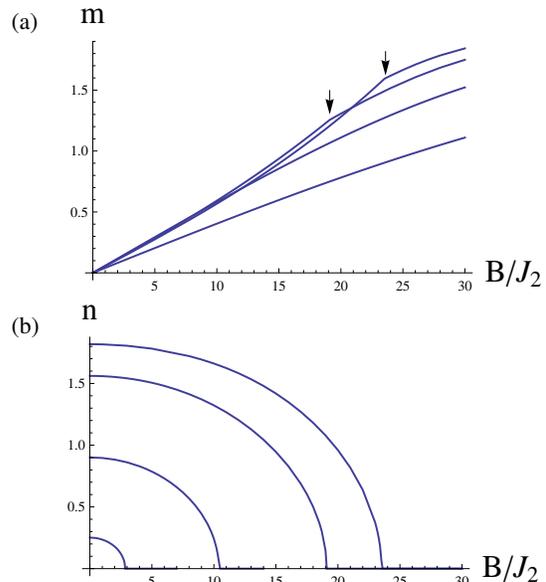
\begin{figure}[tpb]
 \begin{picture}(0,0)
\put(-20.0,+100.0){\textrm{(a)}}
\put(-20.0,-15.0){\textrm{(b)}}
\end{picture}
\includegraphics[width=0.35\textwidth]{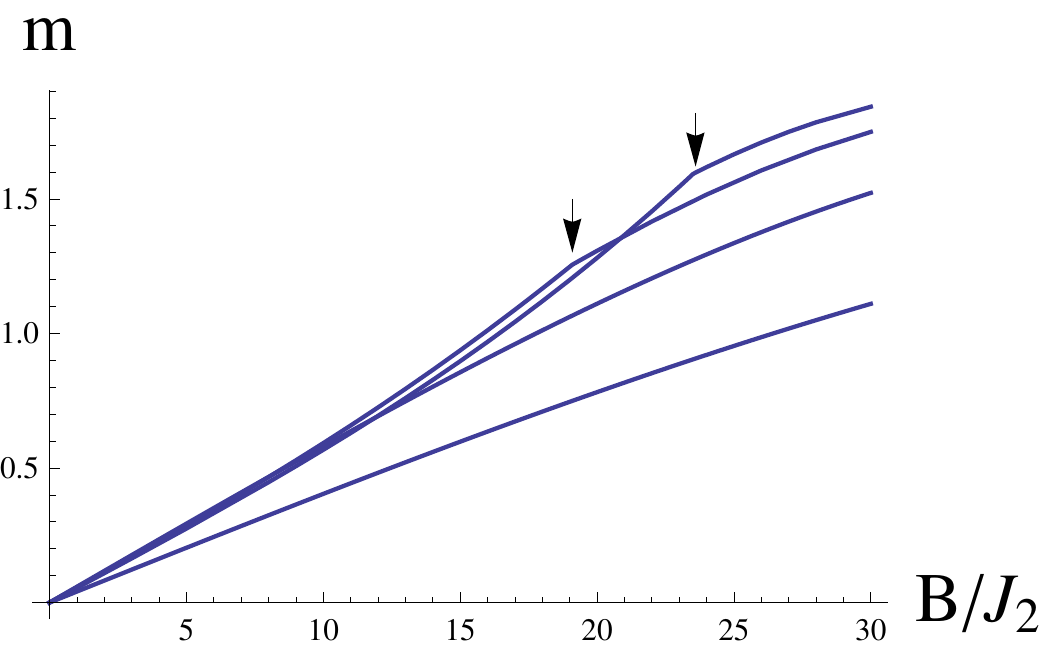}
\vspace{-0.4cm}
\includegraphics[width=0.35\textwidth]{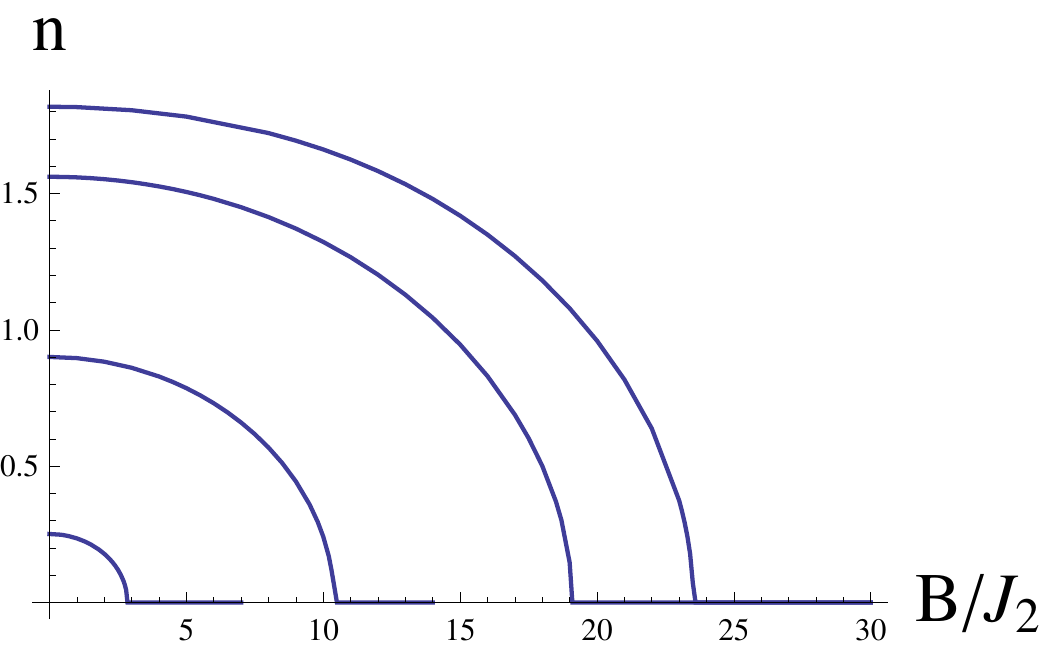}
\caption{(color online).  
(a) Uniform magnetization $m$ versus magnetic field $B$
(in units of the spin exchange $J_2$) for different coupling ratios $x$
(from top to bottom 
$x=0.15,0.10,0.05,0.02$, respectively). The arrows indicate the position of the QCP.
(b) Staggered magnetization $n$ as a function of $B$ (in units of the
spin exchange $J_2$) for $x=0.15,0.10,0.07,0.063$ (from
top to bottom curve).}
\label{fig:magnetizations}
\end{figure}

The minimal $J_2-\lambda$ model in an external magnetic field $B$ 
contains only onsite and NNN interactions. Therefore the diamond lattice
decomposes into two fcc sublattices with $J_2$ playing the role of the NN exchange 
interaction within each sublattice. The Hamiltonian is
\begin{equation}
\begin{split}
\mc{H}_{\text{min}}
&= 
\sum\limits_{\langle ij \rangle} J_2 \ {\bf S}_i \cdot {\bf S}_j  
 + \sum_i \mc{H}^i_0 + B \sum_i S_i^z ,
\end{split}
\label{eq:min_mod}  
\end{equation}
where $\langle ij \rangle$ represents nearest neighbor sites on an fcc sublattice
and $\mc{H}_0^i$ is defined by Eq.~\eqref{eq:ham_parent c}.
In the absence of a magnetic field and when $J_2 / \lambda > x_c = 1/16$
the system is in the CO phase~\cite{chen:prl} with the spins aligned
antiferromagnetically along one of the three cubic axes (here taken to be the $x$ axis).
To decouple the exchange interaction in Eq.~(\ref{eq:min_mod})   we employ mean field theory
with the following ansatz for the average spin at site $i$
\begin{equation}
\langle {\bf S}_i \rangle = n \cos{({\bf p}\cdot {\bf r}_i)}\hat{x} + m \hat{z}
\; .
\label{eq:mft_ans}
\end{equation}
Here, $m$ and $n$ denote uniform and staggered magnetizations, respectively,
and ${\bf r}_i$ are the usual half-integer coordinates of the fcc sites.
In the CO state the spiral momentum 
takes the form ${\bf p} = 2\pi (1,0,0)$ thereby encoding the
antiferromagnetic order along the $x$ axis.
In the disordered SOS phase the staggered moment is vanishing, $n=0$.
With this, the resulting single-site mean field Hamiltonian reads
\begin{subequations} \label{eq:mft_ham}
\begin{equation}
\mc{H}_i^{\text{MF}} = h_z S_i^z + h_x S_i^x+ \mc{H}_0^i
\;,
\end{equation}
where we have introduced the two effective magnetic fields 
\begin{equation}
h_z \equiv 12 J_2 m + B, \ \ \ h_x \equiv - 4 J_2 n \;.
\label{eq:aux_var}
\end{equation}
\end{subequations}
At zero temperature the self-consistent mean-field equations
for the uniform and staggered magnetizations are given by
\begin{subequations} \label{MF eq}
\begin{eqnarray}
m &=& \frac{\partial \epsilon(h_x,h_z)}{\partial h_z},\label{eq:self1} \\
n &=& \frac{\partial \epsilon(h_x,h_z)}{\partial h_x} \label{eq:self2} ,
\end{eqnarray}
\end{subequations} 
where $\epsilon(h_x,h_z)$ denotes the ground state energy of the
mean-field Hamiltonian~\eqref{eq:mft_ham}.  The numerical solutions to
these equations are presented in Fig.~\ref{fig:magnetizations}.  We find
that with increasing field the staggered magnetization is suppressed,
and eventually the magnetic order is destroyed. Hence, the critical
coupling ratio $x_c$ moves from $x_c = 1/16$ to larger values with
increasing field (see Fig.~\ref{fig:ph2}).  The uniform magnetization
shows a small ``shoulder'' at the critical magnetic field when the
staggered magnetization vanishes.


In the neighborhood of the QCP ($|x/x_c-1| \ll 1$)  and for $B$ small compared to $J_2$
it is legitimate to expand the ground state energy $\epsilon (h_x, h_z)$ in the effective magnetic fields $h_x$ and $h_z$.
Up to fourth order we have
\begin{equation}
\begin{split}
\epsilon ( h_x, h_z)
\simeq
- 2 \lambda -  \frac{2}{\lambda } \left(  h_x^2  +  h_z^2 \right) 
+  \frac{2}{\lambda^3} \left(  h^4_x + 4 h^2_x h^2_z +  h_z^4 \right) . 
\\
\end{split}
\end{equation}
Neglecting terms of order $h_z^3$ and higher,
we obtain the following expressions for
the uniform magnetization 
\begin{eqnarray}
m = \left\{
\begin{array}{ll}
-\frac{4B}{48J_2 + \lambda}, & \qquad  \text{SOS} , 
\vspace{0.2cm}
\\
-\frac{(8J_2 - \lambda )B}
{2 J_2 ( 48 J_2  -7\lambda  )}, & \qquad \text{CO} ,
\end{array}
\right.
\;
\label{eq:uniform_mag}
\end{eqnarray}
in the SOS and CO phase, respectively.
Similarly, the staggered magnetization is given by
\begin{eqnarray}
n = \left\{
\begin{array}{ll}
0 ,                    & \qquad \text{SOS} , \\
8\sqrt{2(x-x_c(B))} ,   &\qquad  \text{CO} ,
\end{array}
\right.
\;
\end{eqnarray} 
with
\begin{equation}
x_c\left(B\right) = \frac{1}{16} + \frac{B^2}{16384 J_2^2}
\; ,
\label{eq:qcp}
\end{equation}
 where we retained only the lowest order term in $B$.
The approximate result for the uniform magnetization, Eq.~\eqref{eq:uniform_mag}, describes 
the linear dependence on magnetic field in the regime where $B$ is
small compared to $J_2$ (cf.\ Fig.~\ref{fig:magnetizations}).
It is interesting to note that this behavior agrees with the measured low-temperature 
magnetic susceptibility in \fss.  
Indeed, it is found in Ref.~\onlinecite{loidl:nsnmr} that
the magnetic susceptibility in \fss\ at $T  \to 0$ saturates to a constant value,
independent of $B$.

To conclude, we find that an external magnetic field leads to a
suppression of spin ordering. On the ordered side of the QCP there is a
phase transition to the disordered SOS phase as the magnetic field is
increased.  This behavior is quite different from a {\sl spin} singlet
phase in a typical spin-only system such as
TlCuCl$_3$\cite{nikuni1971bec} or BaCuSi$_2$O$_6$\cite{jaimePRL04}, where only the
magnetic triplet excited states -- magnons -- respond to the magnetic
field, whereas the non-magnetic singlet ground state is unaffected.
Hence, the field stabilizes spin order by Bose-Einstein condensation of
magnons.  Here, however, the strong spin-orbit interaction leads to very
different physics.  Specifically, SOS is not a spin singlet, but a
highly entangled quantum state of spin and orbital degrees of freedom.
As a consequence, it responds strongly to the applied field, and indeed
takes better advantage of the field than does the ordered N\'eel state.

\section{Dynamical spin susceptibility and energy gaps}
\label{sec:subsec22}

In this section, we compute the dynamical spin susceptibility in the SOS
phase with the exchange coupling $J_2$ treated within the RPA.  This
could be compared to inelastic neutron scattering data.  The energy gaps
to the low-lying collective modes are derived from an analysis of the
pole structure of the dynamical spin susceptibility.

\begin{figure}[tb]
\begin{center}
\vspace{-0.3cm}
\includegraphics[width=0.48\textwidth]{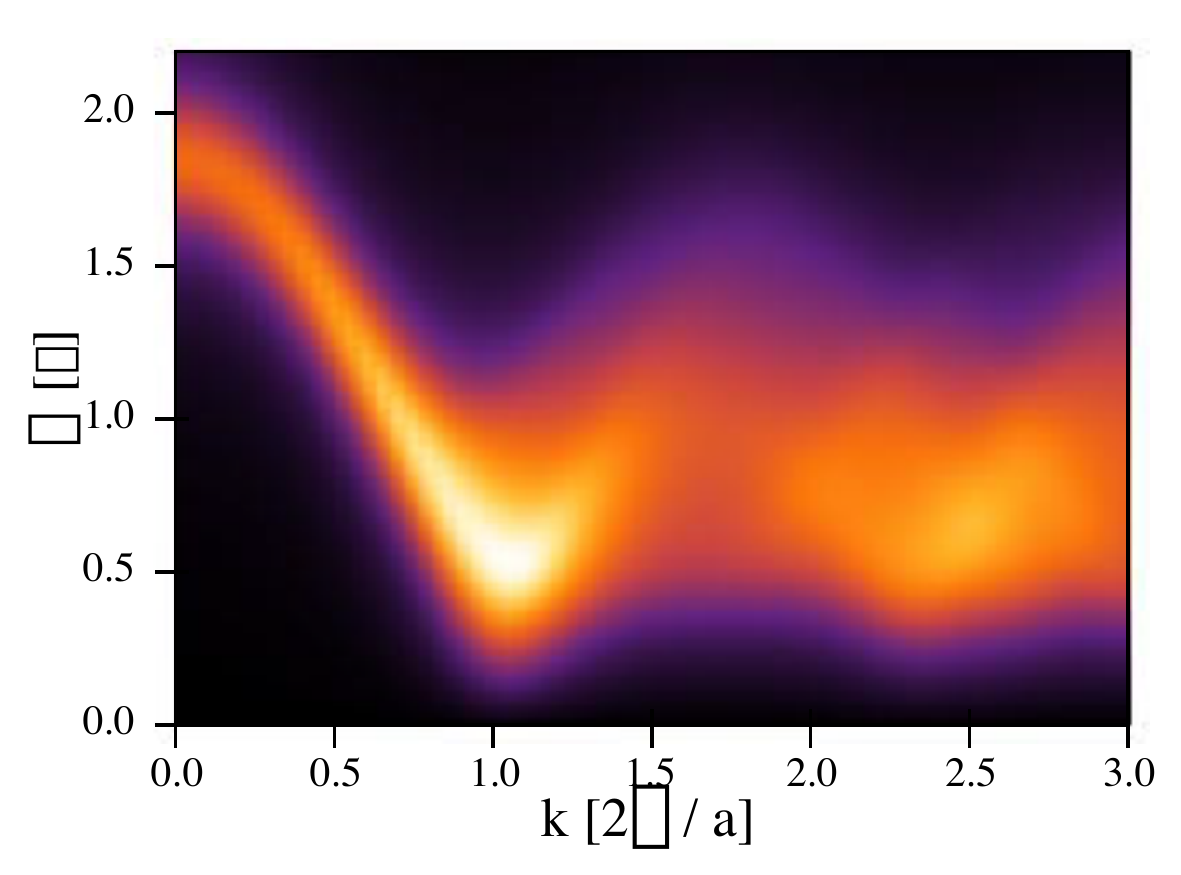}
\end{center}
\vspace{-0.3cm}
\caption{(color online) Imaginary part of the angular-averaged RPA spin susceptibility in zero magnetic field 
$\textrm{Im} \, \chi^{xx}_{\textrm{av}} ( k, \omega; 0 )$
with $J_2 / \lambda = 0.05$ and a damping $\Gamma / \lambda =0.2$.
}
\label{fig:chi 3D}
\end{figure}

In the SOS phase we can approximate the full magnetic susceptibility of the $J_2-\lambda$ 
model $\chi^{\mu \nu} ({\bf k}, \omega ; B)$
by the RPA in terms of the non-interaction susceptibility $\chi^{\mu \nu}_0 (\bm{k}, \omega, B)$
of the on-site Hamiltonian $\mc{H}_0^i + B S^z_i$ [see Eq.~(\ref{eq:ham_parent c})].  
For the $xx$-component we have  
\begin{equation}
\chi^{xx} ({\bf k}, \omega ; B) = \frac{\chi_0^{xx} (\omega)}{1 - \mc{J}({\bf k}) \chi_0^{xx} (\omega) }
\;,
\label{eq:dspin_sus}
\end{equation}
where  $\mc{J}({\bf k})$ is the Fourier transform of the exchange coupling,
\begin{equation}
\mc{J}({\bf k}) = \sum_{\{ {\bf A} \}}   J_2 \cos ( {\bf k}\cdot {\bf A} ) ,
\end{equation}
with $\{{\bf A}\}$ denoting the 12 NNN lattice vectors.
The single-site spin susceptibility $\chi_0^{xx} (\omega; B)$ 
can be constructed from the spectral representation. At zero temperature 
$\chi_0^{xx} (\omega; B) $ is given by
\begin{eqnarray}
\chi_0^{xx} (\omega; B) 
&=&
\sum_{j \neq 0}
\left[
\frac{|\langle 0| S^x  |j \rangle|^2}{ \epsilon_j - \omega - i \Gamma }
+
\frac{|\langle 0| S^x  |j \rangle|^2}{  \epsilon_j  +\omega + i \Gamma }
\right] , \;\; \; \;
\end{eqnarray}
where $| 0 \rangle$ and $| j \rangle$ are the ground state and excited states of the on-site term $\mc{H}_0^i + B S^z_i$, respectively.  The energy difference between the excitated state $| j \rangle$ and the ground state $| 0 \rangle$ is denoted by $\epsilon_j$. Finite lifetime effects are parametrized by a phenomenological damping $\Gamma > 0$. In order to facilitate a direct comparison with neutron scattering data on polycrystalline samples
we perform a numerical average of $\chi^{xx} ({\bf k}, \omega ; B)$ over the angular components of the wavevector $\bm{k}$. For a given wavevector magnitude  $k = | \bm{k} |$ we define 
the angular-averaged spin susceptibility by
\begin{eqnarray}
\chi^{xx}_{\textrm{av}} ( k, \omega; B )
=
\int  \sin \theta \, d  \theta d \phi \,
 \chi^{xx}_{\ } ( \bm{k}, \omega; B ),
\end{eqnarray}
where $\theta$ and $\phi$ describe the direction of the wavevector $\bm{k}$.
Since the inelastic neutron scattering intensity is proportional to the imaginary part
of  $\chi^{xx}_{\textrm{av}} ( k, \omega; B )$ we compute $\textrm{Im} \chi^{xx}_{\textrm{av}} ( k, \omega; B )$
as a function of energy transfer $\omega$ and wavevector magnitude $k$.
Fig.~\ref{fig:chi 3D} displays the numerically calculated dynamical spin susceptibility $\textrm{Im} \chi^{xx}_{\textrm{av}} ( k, \omega; 0 )$ in zero magnetic field. The excitation minima near  $k=2\pi /a$ and $k=5 \pi /a$ agree well with the neutron scattering data on polycrystalline samples.\cite{loidl:ns}

\begin{figure}[t!]
\includegraphics[width=7.3cm]{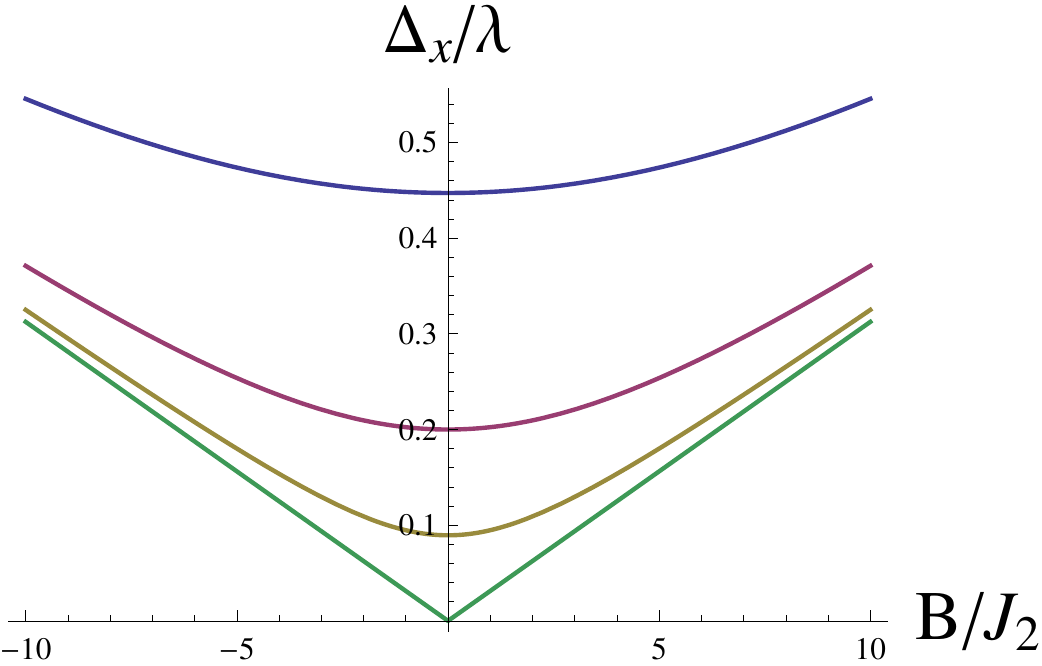}
\caption{(color online). Energy gap $\Delta_x$ (in units of the spin-orbit coupling $\lambda$) 
as a function of magnetic field (in units of the spin-exchange $J_2$) for different coupling rations
 $x$ (from top to bottom $x = 0.05,0.06,0.062,1/16$).}
\label{fig:gapx}
\end{figure}

The dispersing excitation branch shown in Fig.~\ref{fig:chi 3D}  is a collective mode associated with zeros of the real part of the denominator in Eq.~\eqref{eq:dspin_sus}. For the unaveraged dynamical spin susceptibility $\textrm{Im} \chi^{xx}_{\ } (\bm{k}, \omega, B)$ the frequency minimum $\Delta_{x} $ of the dispersing collective mode  occurs at ${\bf k} =2\pi(1,0,0)$. Upon approaching the QCP from the disordered side and for small magnetic field $B$ we find that the gap $\Delta_x$ is vanishing as 
\begin{equation} \label{eq:gap_x}
\Delta_x = 4 \lambda 
\sqrt{  x_c (B) -x  },
\end{equation}
where $x_c(B)$ is defined by Eq.~\eqref{eq:qcp}. 
[The same result also holds for the energy minimum $\Delta_y$  of the collective mode described by $\textrm{Im} \chi^{yy} ({\bf k},\omega;B)$.]
Fig.~\ref{fig:gapx} depicts the gap $\Delta_x$, Eq.~(\ref{eq:gap_x}), as a function of magnetic field for different coupling
ratios $x$ in the SOS phase.


\section{Discussion}
\label{sec:sec5}

\subsection{Summary}
\label{sec:summary}

In this work we have refined the theory, developed in
Ref.~\onlinecite{chen:prl},  of the QCP in a spin-orbital
Hamiltonian for the A-site spinel compound \fss.  The model exhibits an
interesting quantum critical point: on increasing the second neighbor
spin-exchange interaction $J_2$ it passes through a zero temperature
phase transition from a spin-orbital singlet state to a magnetically and
orbitally ordered phase.  First, we considered the effects of a weak
nearest-neighbor exchange interaction $J_1$, which induced a narrow
region of incommensurate phase near the QCP.  We studied the associated
commensurate-incommensurate transition.  Next, we included the effects
of an external magnetic field.  While the quantum critical point studied
here seems
similar to the one found in, e.g., bilayer Heisenberg antiferromagnets, its
behavior under an external magnetic field is quite different.  Namely,
we found that a magnetic field suppresses magnetic and orbital order,
and a transition from the ordered state to the spin-orbital singlet
phase occurs at some critical field strength (see Fig.~\ref{fig:ph2}).
From these findings, we conclude that \fss, which is close to the
quantum critical point, but in the spin-orbital singlet phase, does not
show any field induced transition to an ordered state.  Indeed, recent
NMR experiments in fields up to 8.5T showed no signs of magnetic
ordering.\cite{loidl:nmr}  
Furthermore, we computed the dynamical spin susceptibility in the SOS
phase by means of a random phase approximation.  Averaging our results
over the angular components of the wavevector we performed a comparison
with available neutron scattering data on polycrystalline \fss\ samples
and found reasonable agreement (see Fig.~\ref{fig:chi 3D}).

\subsection{Experiments}
\label{sec:impl-exper}

\subsubsection{Magnetic probes}
\label{sec:magnetic-probes}

The theory expoused in this paper and Ref.~\onlinecite{chen:prl} is broadly
consistent with the results of a variety of magnetic probes applied to
\fss.  It explains the small but non-zero spin gap measured in inelastic
neutron scattering and NMR 1/T$_1$ relaxation rate measurements, as well
as the temperature dependence of the uniform magnetic susceptibility.
The present calculation of the dynamical spin susceptibility matches
reasonably well with experiment.  The lack of field-induced magnetic
ordering is also in agreement with the calculations in this paper.

\subsubsection{Specific heat and disorder}
\label{sec:spec-heat-disord}

The specific heat data on \fss\ reveal several energy scales.  The
magnetic specific heat divided by temperature, $C_m/T$, exhibits a peak
at $T\approx 6K$.  The integral of $C_m/T$ exceeds the spin-only entropy
$R \ln 5$, approaching instead $R (\ln 5 + \ln 2)$ for $T \gtrsim 60K$,
evidencing the 2-fold orbital contribution.  This is quite consistent
with the present model.  However, the lower temperature behavior is more
complex.  For $T<2K$, experiments are fitted approximately by $C_m \sim
A T + B T^{2.5}$, with the linear term dominant for $T< 0.2K$.  The
latter behavior appears at odds with the indications of an energy gap
of $1-2K$ in neutron scattering and NMR experiments.  

A possibility reconciliation of these observations is in the effects of
disorder.  Microscopically, we expect the dominant type of disorder to
be inversion defects, in which the A and B sublattice atoms are
interchanged.  Inversion is very common in spinels.  To understand the
effects of such defects, we apply general arguments based on the Landau
expansion and the theory of disordered systems.  These arguments depend
very little upon the specific nature of the defects, other than that
they are random, not very correlated, and do not break time-reversal
symmetry.

These conditions lead to an important observation: since the
order parameters ${\boldsymbol \psi}_{\mu, a}$ are {\sl odd} under
time-reversal, disorder couples only quadratically to them.  Thus
impurities behave, from the point of view of critical behavior, as
random bonds rather than random fields.  In three dimensions, it is
known that in this case both {\sl phases} are perturbatively stable to
weak impurities.  However, sufficiently close to the QCP, even weak
impurities become non-perturbative.  More formally, random bond disorder
is a {\sl relevant} variable at the QCP.  Physically, the most important
effect of disorder is to locally break the degeneracy of the different
ground states of the clean system within the ordered phase.  For
instance, a specific impurity configuration might favor the ${\bf p}=
2\pi(1,0,0)$ state in one region and ${\bf p}=2\pi(0,1,0)$ state in
another.  Far from the QCP, the surface energy cost to create a domain
wall between the two states overwhelms the random energy gain, and the
system remains uniform.  However, close to the QCP, the surface tension
becomes small, and one expects the system to break into domains.  Thus
impurities induce a non-uniform disordered magnetic state, a ``cluster
spin glass'', near the QCP.  We expect, moreover, that this cluster spin
glass state extends slightly {\sl past} the QCP into the region of the
SOS state of the clean system.  This occurs because the system lowers
its energy slightly more than in the clean case, by taking advantage of
the impurities locally.  

This scenario provides a possible explanation of the specific heat data.
At low temperature, a $T$-linear specific heat is a generic feature of
spin glasses.  It should occur with a small coefficient $A$ when
disorder is weak.  At higher temperature, one recovers approximately the
intrinsic bulk clean behavior, which would be of the form $C_m \sim B
T^3 f(\Delta/k_B T)$, where $\Delta$ is the energy gap, and
$f(\delta)$ is a monotonic scaling function satisfying $f(0)=1$ and
$f(\delta) \sim \delta^{7/2} e^{-\delta}$ for $\delta \gg 1$.  The $T^3$
dependence is characteristic of the linearly-dispersion modes at the
QCP, which is cut off by the gap.  It seems plausible that the
experimental observed $T^{2.5}$ dependence reflects the attempt to fit
such a form to a single power-law.  If the impurities are not too weak,
it is also plausible that they modify the $T^3$ behavior somewhat.  In
any case, the overall behavior seems reasonably in line with theoretical
expectations.

\subsection{Directions for future work}
\label{sec:open-issues}

The theory in this paper (and Ref.~\onlinecite{chen:prl}) appears to give
a consistent explanation for the experimental results on \fss.  However,
there are a number of directions that could be explored in the future.
It would be desirable to have a {\sl direct} proof of the postulated
spin-orbital entanglement in the ground state of \fss.  Theoretical
proposals and experimental studies to this end would be welcome.  Given
the smallness of the gap in \fss, there is a possibility that it might
be driven across the QCP by pressure, which would be very exciting.  
Looking more broadly, it appears that the mechanism for quantum
criticality described here could apply at the very least to any material
with Fe$^{2+}$ ions in a tetrahedral environment.  It would be
interesting to survey such compounds for signs of this physics.

\acknowledgments

This work was supported by the DOE through Basic Energy Sciences grant
DE-FG02-08ER46524. LB's research facilities at the KITP were supported
by the National Science Foundation grant NSF PHY-0551164.

\appendix

\section{Hubbard-Stratonovich transformation and Landau action}
\label{sec:appendix}

In this appendix, we derive the effective Landau action, Eqs.~\eqref{eq:3} and \eqref{eq:free_energy2}, 
from the microscopic Hamiltonian~\eqref{eq:ham_parent}
using the Hubbard-Stratonovich method to decouple the exchange
interactions. In this way, we can relate  the coefficients in Eqs.~\eqref{eq:3} and \eqref{eq:free_energy2} to the 
 microscopic exchange coupling parameters $J_i$  and the spin-orbit coupling $\lambda$.
We consider the $J_1$-$J_2$-$\lambda$ model on the diamond lattice,
\begin{equation}
\mc{H} =  \sum_{\langle ij \rangle} J_1 \ {\bf S}_i \cdot {\bf S}_j 
        + \sum_{\langle\langle ij \rangle\rangle} J_2 \ {\bf S}_i \cdot {\bf S}_j 
        + \sum_i \mc{H}_0^i  
        \; ,
\end{equation}
where the brackets $\langle ij \rangle$ and $\langle \langle ij \rangle \rangle$ denote the summation over first and second nearest neighbors, respectively. 
The on-site spin-orbit coupling term $\mc{H}_0^i  $ is given by Eq.~\eqref{eq:ham_parent c}. 
The partition function reads
\begin{eqnarray}  
{\mathcal Z}  
&=&
{\text{Tr}} \exp{ \left[- \beta \sum_{i,j} J_{ij} \bm{S}_i \cdot \bm{S}_j 
- \beta \sum_i \mc{H}_0^i  \right]}  ,
\end{eqnarray}
with the exchange coupling matrix $J_{ij}$. Here,  $J_{ij} = J_1$ or $J_2$, when $ij$ connects
first neighbor or second neighbor sites, respectively.
We decouple the exchange interaction   by
introducing the auxiliary field $\bm{\phi}_i$ and transforming the partition function to
\begin{eqnarray}
{\mathcal Z} 
&=& 
\int \mc{D} \boldsymbol{\phi} 
e^{   \frac{\beta}{2} J^{-1}_{ij} \boldsymbol{\phi}_i \cdot \boldsymbol{\phi}_j  }    
\ {\text{Tr}} \, e^{ - \beta  \sum_i \left( \mc{H}_0^i + {\bf S}_i \cdot \boldsymbol{\phi}_i \right) }  .
\end{eqnarray}
Expanding around the saddle point yields
\begin{eqnarray}
{\mathcal Z} 
&=& \int \mc{D} \boldsymbol{\phi} \exp{ \left[ -\mc{S}_{\text{eff}} \right] }
\;,    
\end{eqnarray}
with the effective action 
\begin{widetext}
\begin{equation} \label{eq:seff}
{\mathcal S}_{\text{eff}} = \int_0^{\beta} d\tau 
               \left[ -\frac{1}{2} {J}^{-1}_{ij} \boldsymbol{\phi}_i \cdot \boldsymbol{\phi}_j 
               +  \frac{2\partial_{\tau} \phi_i \cdot \partial_{\tau} \phi_i}{\lambda^3}
               -  \frac{2\phi_i \cdot \phi_i}{\lambda} + 2\frac{(\phi_i \cdot \phi_i)^2 +2 (\phi_i^x \phi_i^y)^2
               + 2 (\phi_i^x \phi_i^z)^2 + 2 (\phi_i^y \phi_i^z)^2}{\lambda^3} \right]
               \;.
\end{equation}
Assuming $J_1 \ll J_2$ and expressing the $\bm{\phi}_{\mu}$ field in terms of the staggered
magnetizations $\bm{\psi}_{\mu, a}$
\begin{equation}
\boldsymbol{\phi}_{\mu}({\bf r}_i) 
= (-)^{2x_i} \boldsymbol{\psi}_{\mu,x}+(-)^{2y_i} \boldsymbol{\psi}_{\mu,y}+(-)^{2z_i} \boldsymbol{\psi}_{\mu,z}
\;,
\label{eq:transform}
\end{equation}
we obtain the quadratic part of the free energy density
\begin{eqnarray} \label{eq:f2}
f^{(2)}(\{\psi\})
= \sum_{{\bf k}} \sum_a \left(\frac{1}{8J_2} - \frac{2}{\lambda} 
+ \frac{1}{32J_2}(1+\frac{J_1^2}{4J_2^2}) k_a^2 \right) \big[ |\boldsymbol{\psi}_{A,a} ({\bf k}) |^2+|\boldsymbol{\psi}_{B,a} ({\bf k}) |^2 \big] 
+ \frac{i\ J_1}{16 J_2^2} k_a \boldsymbol{\psi}_{A,a} (- {\bf k}) \cdot \boldsymbol{\psi}_{B,a} ( {\bf k}) .
\label{eq:free_energy3}
\end{eqnarray}
\end{widetext}
We note that Eq.~\eqref{eq:f2} is compatible with 
Eqs.~\eqref{eq:3} and~\eqref{eq:free_energy2},
which we derived using symmetry considerations.
By comparing the coefficients in Eqs.~\eqref{eq:3} and~\eqref{eq:free_energy2}
to those in Eq.~\eqref{eq:f2}
we find the following relations
\begin{eqnarray}
\begin{array}{lll}
v_1^2 &=& 1/(3
2J_2)(1+\frac{J_1^2}{4 J_2^2}), \vspace{0.2cm} \\
r     &=& 1/(8J_2)-2/\lambda, \vspace{0.2cm}  \\
\gamma &=& J_1/(16J_2^2) .
\end{array}
\label{eq:coupling1}
\end{eqnarray}
Similarly, expressing the quartic terms in Eq.~\eqref{eq:seff} in terms
of the staggered magnetizations one can show that the coefficients $g_1$ and $g_2$
are given by
\begin{eqnarray}
\begin{array}{lll}
g_1 &=& 4/\lambda^3  \vspace{0.2cm} ,  \\
g_2 &=& -2/\lambda^3 .
\end{array}
\label{eq:coupling2}
\end{eqnarray}


\bibliography{ref} 

\end{document}